\documentclass[10pt,journal,compsoc]{IEEEtran}
\ifCLASSOPTIONcompsoc
  \usepackage[nocompress]{cite}
\else
  \usepackage{cite}
\fi
\ifCLASSINFOpdf
\else
\fi
\usepackage{booktabs} 
\usepackage{rotating}
\usepackage{amsmath,amssymb,amsfonts}
\usepackage{algorithmic}
\usepackage{array,multirow,graphicx}
\usepackage{float}
\usepackage[inline]{enumitem}

\usepackage{textcomp}
\usepackage{tikz}
\usepackage{adjustbox}
\usepackage{bbding}
\usepackage{graphicx}
\usepackage{epstopdf}
\usepackage{balance}
\usepackage[hyphens,spaces,obeyspaces]{url}
\usepackage{listings}

\newcommand{\subhead}[1]{\par\vspace {1pt}\noindent{\textbf{#1.}}}
\newcommand{\subsubhead}[1]{\par\vspace{0pt}\noindent{\emph{#1.}}}

\lstdefinelanguage{JavaScript}{
	keywords={typeof, new, true, false, catch, function, return, null, catch, switch, var, if, in, while, do, else, case, break},
	keywordstyle=\color{blue}\bfseries,
	ndkeywords={class, export, boolean, throw, implements, import, this},
	ndkeywordstyle=\color{darkgray}\bfseries,
	identifierstyle=\color{black},
	sensitive=false,
	comment=[l]{//},
	morecomment=[s]{/*}{*/},
	commentstyle=\color{purple}\ttfamily,
	stringstyle=\color{red}\ttfamily,
	morestring=[b]',
	morestring=[b]"
}

\begin{document}
\title{Playing With Danger: A Taxonomy and Evaluation of Threats to Smart Toys}

\author{Sharon~Shasha,
	   Moustafa~Mahmoud,
       Mohammad~Mannan,
       and~Amr~Youssef
\thanks{S.~Shasha, M.~Mahmoud, M.~Mannan, and~A.~Youssef are with Concordia Institute for Information Systems Engineering, Concordia University, Montreal, Canada
\protect\\}
\thanks{Copyright (c) 2012 IEEE. Personal use of this material is permitted. However, permission to use this material for any other purposes must be obtained from the IEEE by sending a request to pubs-permissions@ieee.org.}
}

\markboth{IEEE Internet of Things Journal,~Vol.~xx, No.~xx, Month~yyyy}%
{Shell \MakeLowercase{\textit{et al.}}: Bare Demo of IEEEtran.cls for Computer Society Journals}

\newcommand{\barbie}{Hello Barbie}
\newcommand{\toymail}{Toymail}
\newcommand{\sphero}{Sphero BB-8}
\newcommand{\chip}{Wowwee Chip}
\newcommand{\monkey}{Smart Toy Monkey}
\newcommand{\wiggy}{Wiggy Piggy Bank}
\newcommand{\ozobot}{Ozobot Bit}
\newcommand{\cozmo}{Anki Cozmo}
\newcommand{\osmo}{Osmo}
\newcommand{\vtechtab}{VTech InnoTab MAX}
\newcommand{\cloudpets}{Cloudpets}

\newcommand{\pii}{PII collected and transmitted}
\newcommand{\secmes}{Security measures}
\newcommand{\toyapp}{Toy and companion app communication}
\newcommand{\analytics}{Ads and analytics services}
\newcommand{\vulnerabilities}{Potential vulnerabilities}
\newcommand{\coppa}{COPPA compliance}

\newcommand{\local}{No-local-PII-protection}
\newcommand{\play}{Unauthorized-use} 
\newcommand{\configphys}{Unauthorized-config-physical} 
\newcommand{\confignearby}{Unauthorized-config-nearby}
\newcommand{\hotspot}{Unencrypted-hotspot}
\newcommand{\bluetooth}{Insecure-Bluetooth-practice}
\newcommand{\remains}{Always-on}
\newcommand{\rest}{Password-reset-page-does-not-expire}
\newcommand{\bruteforce}{Online-password-bruteforce}
\newcommand{\login}{Login-failure-information}
\newcommand{\remote}{No-remote-PII-protection}
\newcommand{\parental}{Weak-parental-PII-control}
\newcommand{\third}{Exposure-to-third-parties}
\newcommand{\comm}{Unencrypted-comm-channels}
\newcommand{\cookie}{Insecure-session-cookies}
\newcommand{\redirect}{URL-redirect}
\newcommand{\tls}{Insecure-TLS-practices} 

\newcolumntype{R}[2]{%
	>{\adjustbox{angle=#1,lap=\width-(#2)}\bgroup}%
	l%
	<{\egroup}%
}
\newcommand*\rot{\multicolumn{1}{R{60}{1em}}}

\newcommand{\no}{\color{black}{\small {$\times$}}}

\newcommand{\red}[1]{\textcolor{red}{#1}} 
\newcommand{\mm}[1]{\textcolor{red}{MM #1}} 
\newcommand{\modified}[1]{\textcolor{blue}{#1}}

\IEEEtitleabstractindextext{%
\begin{abstract}
Smart toys have captured an increasing share of the toy market, and are growing ubiquitous in households with children. Smart toys are a subset of Internet of Things (IoT) devices, containing sensors, actuators, and/or artificial intelligence capabilities. They frequently have internet connectivity, directly or indirectly through companion apps, and collect information about their users and environments. Recent studies have found security flaws in many smart toys that have led to serious privacy leaks, or allowed tracking a child's physical location. Some well-publicized discoveries of this nature have prompted actions from governments around the world to ban some of these toys. Compared to other IoT devices, smart toys pose unique risks because of their easily-vulnerable user base, and our work is intended to define these risks and assess a subset of toys against them. We provide a classification of threats specific to smart toys in order to unite and complement existing adhoc analyses, and help comprehensive evaluation of other smart toys. Our vulnerability taxonomy addresses the potential security and privacy flaws that can lead to leakage of private information or allow an adversary to control the toy to lure, harm, or distress a child. Using this taxonomy, we perform a thorough experimental analysis of eleven smart toys and their companion apps. Our systematic analysis has uncovered that several current toys still expose children to multiple threats for attackers with physical, nearby, or remote access to the toy.
\end{abstract}

}

\maketitle

\IEEEdisplaynontitleabstractindextext

%
\IEEEpeerreviewmaketitle

\IEEEraisesectionheading{\section{Introduction}\label{sec:introduction}}

\IEEEPARstart{W}{ith} smart toys set to grow to an \$18 billion market by 2023 (as estimated by Juniper Research), we set out to classify the security and privacy threats they pose. Smart toys, which have sensors and/or actuators and internet connectivity, are in effect a subset of IoT. While threats posed by IoT have been extensively studied, we argue that smart toys present a distinctive attack surface because of their unique target user group (children and teenagers). A growing body of research shows that children harbor trust and attachment for smart toys~\cite{breazeal2018intelligence,breazeal2018relationships,breazeal2018conformity,vollmer2018conformity}, and the effects of bugs and exploits in these toys are significantly amplified due to such a vulnerable user base. 

A child's trust for her/his toy can be exploited in several ways. A beloved toy may accompany a child wherever she goes, allowing comprehensive location tracking, e.g., via GPS or IP geolocation~\cite{mcreynolds2017toys}. Bluetooth or WiFi enabled toys may broadcast a static MAC address, further facilitating location tracking. Children are inclined to trust their toys and follow their lead~\cite{turkle2006encounters, vlahos2015barbie}. Our tests on the \sphero\ showed that an unauthorized person could control the movement of the toy, potentially leading a child to follow the toy into danger. Using \wiggy, an unauthorized person could issue rogue instructions to the child, for instance directing her outside the home. In both scenarios, the adversary could be within the Bluetooth range, about 40 meters. Finally, children, who tend to have wider trust boundaries than adults, are more inclined to share confidential and identifying information with a toy, and less likely to understand how that information will be shared and the consequences of sharing~\cite{mcreynolds2017toys,leite2016robot}. Some toys (e.g., Hello Barbie) actively solicit personal information, but many toys (including all the toys we examined) passively gather information, e.g., to infer usage patterns or improve personalization. Toys that require creation of an account online may collect additional information, e.g., email address. Some toys are equipped with microphone or cameras to capture audio, photo, and video---troubling enough if leaked, but in the worst case scenario pose the risk of being turned into an espionage device if an adversary can turn them on at will~\cite{cloherty2017hackable}.

As a vulnerable population, children are at risk of being harmed by smart toy exploits. A child may be psychologically scarred if a trusted toy suddenly starts presenting disturbing, violent, or pornographic content. For instance, some of the toys we examined present content fetched over an unencrypted channel, making them subject to being maliciously replaced by a network attacker. A compromised toy with a speaker can be made to voice distressing audio~\cite{valente2017security}. Finally, a child can be harmed by any leak of information captured by a smart toy. The harm may be psychological; for instance, a child may experience a trusted toy revealing information thought to be confidential as a personal betrayal~\cite{jones2016can}. More saliently, it can result in immediate personal danger, e.g., if it reveals the child's location to an attacker, whether directly or indirectly, while others, like the disclosure of a child's birthday (requested by some of the toys we examined), can facilitate identity theft, which may become apparent only much later when the child reaches adulthood and the consequences are more difficult to undo~\cite{newman2005identity, morris2015vtech}.

Smart toys also share the vulnerabilities inherent to other IoT devices. Among these are the use of companion apps and third party libraries, increasing a toy's attack surface. An insecure or over-privileged companion app may be exploited to access the smartphone's microphone, camera, or GPS~\cite{mylonas2011smartphone}, or directly control the toy; any communication between the toy and the app, if unencrypted, can also be intercepted or modified. Toy makers may store personal data insecurely or for an indefinite duration, increasing the likelihood of a breach, as does the use of third-party ads and analytics libraries~\cite{grace2012unsafe}. Smart toy makers are also often simply the makers of other general purpose toys, with no particular expertise in information technology, particularly information security. They may thus not fully understand potential security risks to address them throughout the product life cycle~\cite{peppet2014regulating}. This includes development, design, and testing phases, as well as after-market care, e.g., secure firmware updates should a vulnerability be discovered~\cite{ftc2015iotreport}.

Several classification schemes have been proposed for general IoT security~\cite{zhao2013survey,roman2013features,abomhara2015cyber,denning2013computer,hunt2017seven}, however nothing specific to smart toys. Some systematic analyses have been conducted into compliance of children's apps with children's privacy laws~\cite{liu2016identifying,reyes2018examining}, and experimental work on selected smart toys specifically~\cite{chu2018security}. Some inroads have been made into developing privacy paradigms for children's online activity and even into smart toys~\cite{Rafferty2017privacy,toys2017stast,yankson2017privacy}. However, no taxonomy has been proposed that systematically addresses both the security and privacy of smart toys, specifically, for experimental evaluation. In this work, we examine eleven Internet-connected toys on the market and determine what Personally Identifiable Information (PII) they collect, what PII is shared with third parties like ads and analytics servers, what steps toy makers take to secure this data, and if the toy can be compromised in a way that poses a threat to the physical or psychological safety of the child using it. We also define a list of attacks categorized by attacker proximity and assess the smart toys against them.

Traditional threat taxonomies devised for general IoT devices do not, in our opinion, adequately address the threats they pose to children specifically. Some threats we present are specific to smart toys (e.g. parental PII protection), while others are given more emphasis (e.g. exposure to third party ads and analytics services) compared to existing taxonomies. Regulatory frameworks, like COPPA in the US, that address children's digital privacy specifically recognize that children face unique harms from such devices, and are designed to protect children against these threats.

We consider two broad categories of risk: breach of PII, and unauthorized control of the toy. While there is no universal definition of PII~\cite{schwartz2011pii}, we have closely adhered to that specified in COPPA, as the most widely applicable regulation to specifically address children's privacy~\cite{ftcfaq}. We have categorized PII into the following: personal information including name, gender, physical address, email address, telephone number, voice recordings, and photos; device information, including mobile device unique identifiers including IMEI, serial number, MAC address, IP address, and smart toy ID; and service usage information, including session start and end time, session duration, and app features. 
Our findings show that popular smart toys are vulnerable to several attacks in our taxonomy, some of them severely. For example, in \toymail, we found that blocked contacts still receive children's personal information. \wiggy\ allows nearby adversaries to issue arbitrary tasks to a child, in the familiar voice of the toy. \monkey\ sends personal information unencrypted over HTTP to a server located in China. \vtechtab\ does not require any authentication to access a child's personal information.

\subhead{Contributions}
\begin{enumerate}
	\item We develop an experimental framework for evaluating security and privacy of smart toys. The framework encompasses PII collected and transmitted by smart toys, and security measures that have been taken to protect them; and PII collected and transmitted to third parties such as ads and analytics services, and the third parties' TLS practices. Our framework relies on analysis  of network traffic (WiFi and Bluetooth), and reverse engineering and code analysis of companion apps. 
	\item  We investigate a representative, diverse set of eleven smart toys using our experimental setup to expose potential PII leakage, weak security measures, and other vulnerabilities. Our methodology and tests can be directly used for other toys on the market, and extended/adapted for evaluating future toys.  
	\item We uncover, through experimentation, an excessive collection of unique IDs that facilitate tracking children across  different services or platforms, and sending children's PII to unauthorized entities. Our results also show that several toys expose children to threats that may be exploited by physical, nearby, or remote attackers.
\end{enumerate}

As part of responsible disclosure, we have contacted the manufacturers of all toys mentioned in this paper and shared our findings with them. \wiggy\ responded quickly and mentioned that they are looking into the vulnerabilities and possible fixes. \barbie\ responded that their technical team may get back to us ``if they have any concerns.'' \cozmo\ discarded our findings, describing them as ``unsolicited ideas.'' We have also received automatic replies from \osmo, \cloudpets, \sphero, and \toymail; the rest have not responded even after 6 months. Overall, most responses (and lack thereof) indicate that these companies take privacy matters much more frivolously than expected.

\vspace{-10pt}
\section{Related work}\label{sec:related}

Smart electronics for children have made the news for security breaches numerous times in the past; e.g., see  Rapid7~\cite{rapid72015iot} (strangers could hack baby monitors to view children sleeping and even talk to them directly), Norwegian Consumer Council~\cite{bouvet2016toys} (highlight dangerous security vulnerabilities in 3 toys: i-Que, My Friend Cayla, and \barbie), Motherboard~\cite{motherboard2016gps} (strangers could track children's location and message them through the HereO smart watch), and BBC~\cite{bbcgermany2017cayla} (German ban on Cayla in February 2017, branding it as an ``illegal espionage apparatus'').  Like wearables for adults, many smart toys are Bluetooth-enabled and accompany children outside the home, and previous research on fitness trackers has uncovered numerous vulnerabilities such as location tracking~\cite{Hilts2016fitness}.

Several security breaches involving the \vtechtab\ have been reported. 
In 2015, VTech's Learning Lodge database, used by the VTech suite of children's tablets, suffered a severe breach, exposing the PII of 6.3 million children's profiles~\cite{vtech2017breach}. The breach was due to a combination of insecure practices, e.g., PII was sent in plaintext (no HTTPS), the database was vulnerable to a simple SQL injection attack, and account passwords were stored using an MD5 hash (no salting). The breach ultimately led to a \$650,000 settlement with the U.S. Federal Trade Commission for unauthorized collection of children's personal information and for failing to secure that information~\cite{ftc2018vtechsettlement}, and a related finding from Canada's Privacy Commissioner that VTech was in breach of its privacy laws (but no corresponding settlement)~\cite{pipeda2018vtech}. Also in 2016, the UK-based security firm Pen Test Partners found that the \vtechtab\ is vulnerable to trivial data extraction~\cite{ptpvtech2015extraction}. (We found that this flaw has not been fixed.) Later in 2016, it was found that pornography could be easily accessed through the built-in browser by using Google Translate, formerly white-listed but since removed by VTech~\cite{yahoo2016vtech}. Rapid7 also found severe vulnerabilities in the Fisher Price Smart Monkey, since fixed~\cite{rapid72016hereo}.

\barbie\ has been extensively scrutinized for its privacy and security practices~\cite{hung2016glance, fantinato2017survey, nbc2015barbie}. It encourages children to divulge intimate details, which are then shared with Mattel's partner, ToyTalk, and with parents through the web portal~\cite{taylor2016smart, jones2016can}, presenting legal implications if a child discloses any physical or sexual abuse to the toy~\cite{moini2016protecting}. In January 2016, Somerset Recon Inc.~\cite{Somerset2016barbie} reported several vulnerabilities in \barbie, some of which are still valid as we found (e.g., broadcasting an open hotspot and allowing unauthorized configuration during pairing). 
Popular brands of children's smart watches have been analyzed by the Norwegian Consumer Council, highlighting vulnerabilities in all watches they examined~\cite{watchout}. In November 2017, ``Which? UK'' issued a report examining common Bluetooth vulnerabilities in children's toys that allowed them to be taken over by nearby attackers~\cite{which2017toyalert}. 
Given the stakes at risk, the US Senate tabled a report on smart toy security in December 2016~\cite{senate2016toys}. The following July, the FBI issued an alert warning parents against privacy and security concerns regarding smart toys~\cite{fbi2017warning}.

Rafferty et al.~\cite{Rafferty2017privacy} argue that the traditional access control model is insufficient to protect the privacy of children using smart toys. They introduce a conceptual model for smart toys, allowing parents to configure privacy rules and receive notifications about sensitive data disclosure. The model assumes that children are oblivious to privacy concerns, and as a result, cannot adequately protect themselves online or anticipate the consequences of leaking private information to smart toy makers and other third parties.

Mahmoud et al.~\cite{toys2017stast} propose an analytical framework with 17 privacy-sensitive criteria to evaluate smart toy privacy practices. They analyzed the available privacy policies and terms-of-use documents for 11 smart toys, and reported potentially excessive collection of children's private information and lack of information on data storage location and legal compliance. They also performed static analysis of companion apps of these toys, and found most apps to be over-privileged with the potential to leak children's private information. Our taxonomy, in contrast, offers a classification of smart toy-specific vulnerabilities. We further investigate the \emph{actual} (as opposed to the \emph{potential}) security and privacy practices of a similar set of smart toys through dynamic analysis by using the toys and apps in a realistic scenario; our experiments measure PII collected by the toy manufacturers and third-party servers, and uncover concrete security practices and vulnerabilities. 

In summary, past work mostly involved isolated case studies exposing vulnerabilities in an individual smart toy, or analytical frameworks for smart toy privacy practices. In this work we propose a taxonomy of vulnerabilities pertinent to smart toys, categorize them by attacker proximity, and comprehensively assess a set of toys against it.

\vspace{-10pt}
\section{Existing regulatory frameworks}\label{sec:regulatory}
Internet-connected toys must comply with traditional toy safety regulations, and as online services, they must adhere to applicable digital privacy laws. Where there are such laws, they tend to be more restrictive with respect to children, in keeping with the common understanding of children requiring greater legal protection. In what follows, we briefly discuss EU and US laws and regulations, as we also evaluate the toys in terms of following relevant regulatory restrictions.

In the EU, the General Data Protection Regulation (GDPR) governs how consent should be obtained when collecting children's personal information. Specifically, a consent must be obtained by a parent or guardian for children under the age of consent (13--16 years, varies among the member states), and privacy notices should be written in clear, age appropriate language~\cite{gdpr-article8children}. Data collected from children is otherwise subject to the same rules that apply to any other EU citizen, including data transparency (the ability of an individual to determine what data has been collected about them), data erasure (the right to erase that data, otherwise known as the right to be forgotten), and data portability (the ability to obtain the data for reuse across online services). The GDPR also mandates how companies must respond in the event of a breach, and penalties may be imposed for failing to report the breach to the applicable authority and notifying the affected individuals in a timely fashion. EU citizens do not have to reside in the EU to be protected by the GDPR, even if they use services from a company without any legal presence in the EU market~\cite{gdpr}.

In the US, the Children's Online Privacy Protection Act (COPPA) governs data collection and online tracking of children under the age of 13. Online services should take reasonable measures to obtain parental consent to collect and disclose information, clearly outline data retention policies, and provide a means for data review and deletion. Under COPPA, private information is defined to include persistent identifiers like an IP address, which by default is included in all TCP/IP traffic; thus, by definition, all internet-connected toys collect private information. MAC addresses, like those of a toy's Bluetooth or WiFi interface, are also considered to be persistent identifiers; in our tests, we have observed MAC addresses transmitted to first party toy servers as well as third-party analytics servers.  However, merely collecting persistent identifiers (absent parental consent) is not in and of itself a violation of COPPA, if it is not used in conjunction with any other identifying information, or  if it is not stored or made public, or if it not used for any other purpose than to ensure proper operation of the service (in this case, the toy). If the toy or companion app contacts a third-party analytics company, which then uses persistent identifiers to build a profile of the child, or passively track the child online, it is a COPPA violation, absent explicit parental consent.
Further, the toy's privacy policy must explicitly list all such third party companies~\cite{ftccoppafullact}.

Prior work has analyzed toys' privacy policies for compliance with these laws, and compared, through static analysis, the toys' stated policies with the privileges requested by their companion apps. Through dynamic analysis, we have found excessive collection of persistent identifiers by toy makers and ads and analytics services, and in some cases, we also have found a clear violation of COPPA’s injunction against persistent tracking for the purpose of advertising. In addition, while both COPPA and GDPR do not explicitly address issues of security, we have found security vulnerabilities that expose PII to unauthorized parties (e.g. unauthorized configuration and use), which violates the spirit of such regulations if not their actual content.

\section{Analysis framework}\label{sec:framework}
\begin{figure}[h]
	\centering
	\includegraphics[scale=0.3]{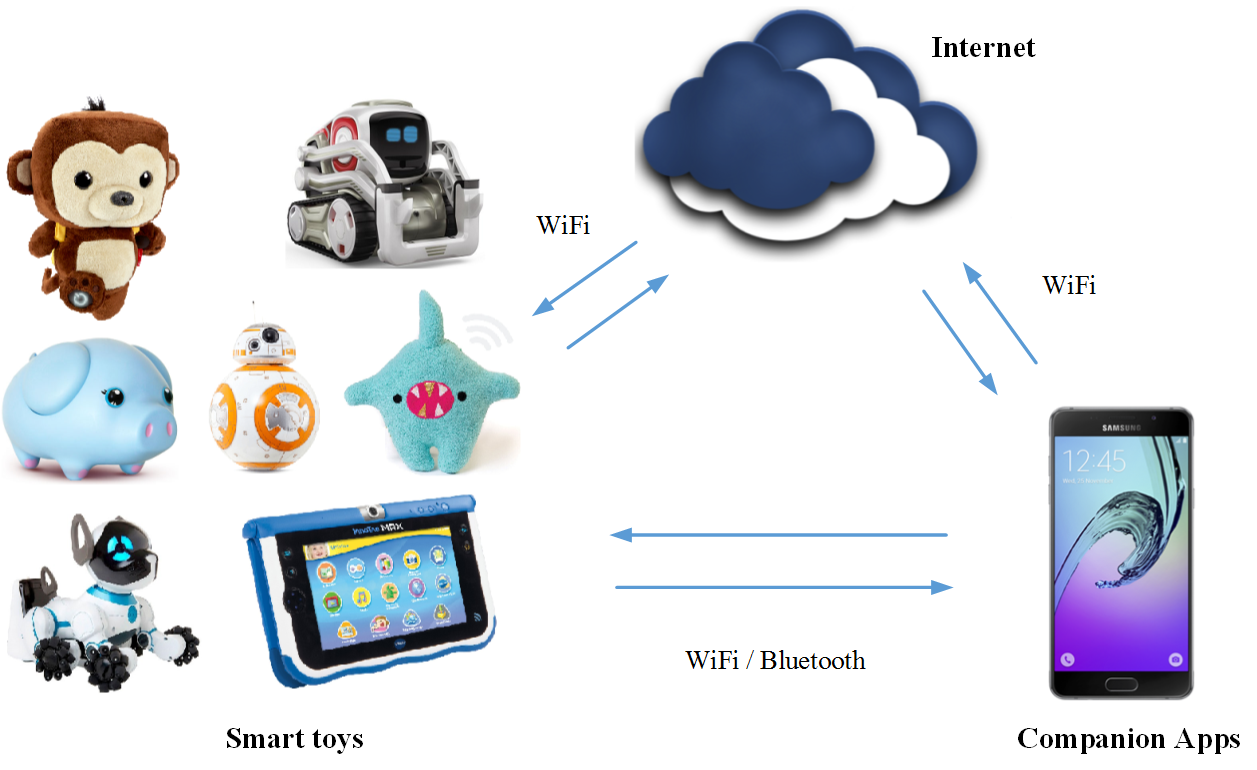}
	\caption{Smart toy attack surface}
	\label{fig:toys_attack_surface}
\end{figure}

Most of the smart toys we analyze are intended to be used with companion apps running on smartphones or tablets. All the mobile apps we examine, and all toys with direct connectivity,  communicate with remote hosts. This presents a wide attack surface covering toys, apps, remote hosts, and the communication between each pair of these; see Fig.~\ref{fig:toys_attack_surface}.

\subsection{Security of PII collection and transmission}

Toys that require the user to create an account typically collect and store more PII than toys that do not. We audit the network traffic of smart toys and their companion apps to determine the PII they collect, transmit, and share with toy servers and third parties, and whether those third parties are ads or analytics servers. 

We evaluate the security measures for protecting the collected PII in local storage and when being sent to a remote server. In some cases, we also verify that non-PII or non-confidential content such as toy name and images (in toy/web UI) are protected from malicious tampering.
Data transmission (including PII) can occur between the toy and a companion app, a companion app and a remote server, or, for toys with Internet connectivity, the toy itself and a remote server. Additionally, data may be stored locally on the toy or on the device hosting the companion app. To investigate the confidentiality and integrity of data, we assess the security of communications channels between each pair of endpoints and local data storage mechanisms.

We determine the protocols used to communicate with remote hosts (e.g., HTTP or HTTPS), 
and assess the likelihood and severity of active man-in-the-middle attacks, from replacing UI elements in the app with inappropriate content to remotely controlling the toy.

\subsection{\vulnerabilities} 
We define a list of potential vulnerabilities and assess the smart toys against them. These vulnerabilities are grouped by the proximity required to exploit them.

\subhead{Physical access}
The following vulnerabilities can be exploited by an adversary with physical access to the toy or companion smart device.
\subsubhead{\local} An adversary can retrieve PII stored locally in the toy's internal storage, or within the companion app, leaving it unprotected in the event the toy/device is lost, stolen, or sold unsanitized.
\subsubhead{\configphys} An adversary can configure the toy/device to maliciously forward PII to their account, or issue harmful commands to the child. 
\subhead{Nearby access} 
These vulnerabilities require an attacker to be near the toy (e.g., within WiFi/Bluetooth range). 
\subsubhead{\play} This vulnerability allows an attacker to use the toy, for instance to remotely control it.
\subsubhead{\confignearby} An attacker can download the companion app in his own device, then connect to the toy, and maliciously configure it. 
\subsubhead{\hotspot} Some toys can toggle their WiFi adapter into access point mode to allow a companion app to connect directly to the toy and configure it. A toy's hotspot should authenticate devices that attempt to connect, and use an encrypted channel. Otherwise, an adversary who connects to a toy's open hotspot can launch an MitM attack and sniff PII in plaintext, or maliciously configure the toy. 
\subsubhead{\bluetooth} The use of static Bluetooth MAC addresses allows a toy, and the child using it, to be tracked persistently (cf.~\cite{Hilts2016fitness}). Also, accepting unauthorized Bluetooth connections by the toy allows adversaries to connect to the toy and change its behavior, or launch an MitM attack to sniff information transmitted between the toy and the app (cf.~\cite{Bluetoothsecurity}). Bluetooth Low Energy (BLE) resolves these two problems by MAC address randomization and whitelisting~\cite{fawaz2016bluetooth}. 
\subsubhead{\remains} Toys that do not have power switches and include sensitive sensors or are connected to the Internet, increase their exposure to potential attack, for instance by continuously broadcasting a static MAC address.

\subhead{Remote access}
The following vulnerabilities expose the toy to remote attacks.
\subsubhead{\bruteforce} Some toys allow Web login to the parent account. Login pages that allow unlimited number of password trials could allow an adversary to brute force passwords, particularly if they are weak.
\subsubhead{\remote} This vulnerability allows adversaries to access PII remotely without proper access control (e.g., allowing access to audio/video recordings through a link, without any further authentication).
\subsubhead{\parental} Some toys do not allow parents to review or delete PII collected from their children, such as voice recordings or pictures.
\subsubhead{\third} Third parties with whom PII is shared, such as ads and anaytics server, may have inadequate measures to protect collected data.
\subsubhead{\comm} Information exchange between different parties (toy, app, and hosts) may be susceptible to interception through a MitM attack (e.g., if HTTP is used).
\subsubhead{\cookie} Session cookies are used to automatically log users in to their accounts. If they are not adequately secured, for instance if they do not expire or the \emph{secure flag} is not set, an adversary may gain unauthorized access to the parent account and to the child's profile. 

\subsubhead{\tls}
TLS vulnerabilities~\cite{xiong2014survey} may result from using weak cipher suites, old TLS versions such as SSL 3.0, and vulnerable extensions. We examine servers contacted by the toy against protocol vulnerabilities such as POODLE, 
CRIME, Heartbleed, and Ticketbleed~\cite{ristic2013bulletproof}, and assess server certificates for security issues, including using weak cryptographic keys and certificate mismatch.
\subsubhead{\redirect} Web servers that are not hardened against a URL redirect vulnerability could allow an adversary using social engineering techniques to redirect users to phishing websites (e.g., to steal their credentials). An adversary could send links to users that appear legitimate (e.g., containing the correct domain name), but that use special characters to redirect the user to a malicious domain. Servers should prevent URL redirect, or at least whitelist accepted URLs. 
It should be noted that an ideal taxonomy of threats would include other common web-based vulnerabilities, such as those in the OWASP top-ten most critical web application security risks. We have not presented these here as they require more intrusive probing of the web sites in question (e.g., cross-site scripting (XSS)), which for legal reasons, require consent from the companies in question.

\section{Methodology and experimental setup}\label{sec:setup}

Using our threat classification model, we perform case studies on eleven well-known smart toys, described in Appendix~\ref{sec:smarttoys}. We systematically assess the toys, their companion apps, and the remote servers they communicate with against our model. To obtain data for our analysis, we systematically run the toys through typical use case scenarios, using real mobile devices to circumvent evasion techniques apps might use to avoid detection of suspicious activity. Existing app analysis tools are used to detect the servers the toys communicate with, the PII they transmit, and the security measures applied by toy makers to protect personal information. We augment these wherever suitable by applying reverse engineering mechanisms, network traffic analysis, retrieval of certificate private keys, and leaking protection passphrases.

\subsection{Experimental setup} 
We maintain two experimental setups. Our first environment uses a PC hosting Windows 10 professional 64-bit. The PC is configured to use a MediaTek 802.11G WiFi adapter in access point mode, using the 2.4 GHz frequency band (the band used by the toys we examine), and a second NIC with Internet connectivity. We examine the companion apps using a Samsung N7100 hosting Android 4.4.2. This version of Android is ranked among the top four Android versions used in the wild~\cite{androidwild}. As an older version of Android, it is illustrative of how toys deal with older cipher suites, TLS versions and TLS protocol vulnerabilities. Our second environment uses a MacBook tethered via USB to an iPhone providing Internet connectivity and configured to act as an access point. Companion apps ran on iOS 10, both on an iPad Air and an iPhone 7. We chose iOS 10 as it was the latest iOS version at the time we began experimentation in June 2017, and statistics show that the majority of iOS users run the latest version of the OS~\cite{iOSVsAndroidVersions}. We tested one group of toys on an Android platform (Toymail, Wiggy,  Barbie, CHiP, and Cloudpets) and the other group on iOS (Sphero, Cozmo, Ozobot, Osmo, and Monkey). We hope to expand on this work in the future by performing a systematic comparison of the latest versions of both platforms.

\subsection{Network analysis}
Both experimental setups use Wireshark\footnote{https://www.wireshark.org} to sniff network traffic initiated by the toys and their companion apps.
To decrypt traffic, our first setup uses Burp Suite\footnote{https://portswigger.net/burp} as a MitM proxy and adds the Burp Suite CA certificate to the smartphone CA store. Our second setup uses MitMProxy\footnote{https://mitmproxy.org} in lieu of Burp Suite, and correspondingly installs MitM-generated certificates on the devices running the apps. We address special cases where this approach was insufficient to intercept and decrypt TLS communications. All such interventions were performed on Android-based client apps that were reverse-engineered. 
The Qualys SSL Labs\footnote{https://www.ssllabs.com} testing suite was used to perform TLS analysis on the server side, and we examined packet captures in Wireshark to analyze TLS practices on the client app.

\subsection{Ads and analytics analysis}
Wherever possible, we intercept data between the companion app and third party ads and analytics servers (Internet-capable toys we examined do not directly communicate with ads and analytics services). To determine the extent to which this data can be used to uniquely identify the user across multiple apps on the same device or across different platforms on different devices, we created an account on a selected analytics server, Flurry Analytics, used by two of our toys, and wrote and deployed a dummy app to connect to it. Using this method, we were able to determine that Flurry Analytics used the same unique identifier on both a companion app and our dummy app, despite the fact that the vendor ID is different on both. This identifier can be used to distinguish the user across unrelated services on the same device. Other analytics servers we examined correlate unique identifiers with users' personal information such as email address, or use cross-platform cookies such as DSID and IDE~\cite{googlecookies} cookies to distinguish users across different platforms.

\subsection{App analysis} \label{subsec:appanalysis}
All companion apps were analyzed manually (as opposed to using, for example, the Monkey automation tool\footnote{https://developer.android.com/studio/test/monkey.html}). We mimicked common use case scenarios with the goal of triggering app UI events looking for signs of PII leakage, weak security measures, or potential vulnerabilities, and captured data from the interaction with the toy and its companion app.

\subhead{Modify custom CA store}
When the companion app uses a custom CA store to verify server certificates, we apply the following methodology to force the app to accept the MitM certificate: (a) Decompile the app using Apktool\footnote{https://github.com/iBotPeaches/Apktool} to retrieve the CA store from the app's assets directory, (b) Patch parts of the app smali files to force revealing the CA store password, (c) Use the password to access and update the custom CA store, (d) Use the keystore explorer tool to add the MitM certificate, and 
(e) Replace the CA store in the assets directory of the app with the new store, rebuild the app using Apktool, re-sign and verify the patched apk file, and use the adb tool to reinstall it. As a result, the app accepts the MitM certificate. 

\subhead{Exfiltrate client SSL certificate}
In cases where the companion app authenticates to the server using a client Bouncy Castle file (PKCS\#12), which encompasses the client's public key and the client's private key, we reverse engineer the app to exfiltrate the PKCS\#12 file and the passphrase used to protect it. We then add the PKCS\#12 file (the certificate and the private key) to the interception proxy. This allows the interception proxy to authenticate to the server. 

\subhead{Bypass certificate pinning}
In cases where the companion app uses certificate pinning to refuse server certificates signed by any CA other than the one with the pinned certificate in the app, we patch the corresponding parts of the app smali files to disable this feature and intercept the communication.

\subsection{TLS vulnerabilities analysis}
We assess toys, apps and server SSL practice to determine potential TLS protocol vulnerabilities, and use Qualys SSL Labs to determine TLS server parameters, including supported TLS versions, cipher suites and extensions. 

\subsection{Bluetooth analysis}

BLEscanner\footnote{\url{https://play.google.com/store/apps/details?id=com.macdom.ble.blescanner}} is used to examine the Bluetooth connection between the toy and the companion app, and determine whether the toy's Bluetooth MAC address is persistent or dynamically changing. We investigate how an adversary could tamper with the communication or gain unauthorized access to Bluetooth parameters to conduct MitM or other attacks~\cite{btmitm}. The BTLEjuice tool\footnote{\url{https://github.com/DigitalSecurity/btlejuice}} is used to conduct MitM attacks on toys that use Bluetooth to exchange data with a companion app, to both sniff and alter the communication, and RamBLE\footnote{\url{https://www.contextis.com/resources/tools/ramble-ble-app}} is used to modify BLE settings.

\section{Summary of Results}\label{sec:results}

We systematically applied our threat classification scheme to a sample of 11 popular smart toys on the market, varying by target age range and functionality. In particular, the toys we selected span multiple target ages, from \barbie\ and \monkey, geared towards younger audiences, to what are generally thought of as STEM (Science, Technology, Engineering, and Math) toys aimed at children 8+, such as \sphero, \cozmo, and \ozobot. Toy functionality encompasses AI  capability,\footnote{For example, \cozmo\ is marketed as an AI toy: ``Anki Cozmo, is an AI toy robot with a big brain and even bigger personality'' (\url{https://www.anki.com/en-ca/cozmo} (in the page metadata).} such as voice and image recognition; sensors, like microphone and camera; mobility; and wireless communication, like WiFi and Bluetooth. Detailed results and analysis of individual toys are available in~\ref{sec:smarttoys}.

Nine of the eleven toys we examined are vulnerable to some form of attacks, whether through physical or nearby access (within WiFi or Bluetooth coverage), or remotely, such as over HTTP; see Table~\ref{tab:attacks}.

\begin{table*}[!htb]
	\small
	\centering
	\caption{Attacks by proximity}
	\label{tab:attacks}
	\begin{adjustbox}{max width= 14.5cm}
	\begin{tabular}{lccccccccccccccccccc}
		&&&\multicolumn{3}{p{1.5cm}}{\centering Physical}&&\multicolumn{5}{p{2.5cm}}{\centering Nearby }&&\multicolumn{7}{p{3.5cm}}{\centering Remote}\\
		\cline{4-6}  \cline{8-12}  \cline{14-20} 
		~  &
		\rot{\local } &
		\rot{\configphys} & &
		\rot{\play} &
		\rot{\confignearby} &
		\rot{\hotspot } &
		\rot{\bluetooth} &			
		\rot{\remains } & &
		\rot{\bruteforce} &
		\rot{\remote} &
		\rot{\parental} &
		\rot{\third} &
		\rot{\comm} &
		\rot{\cookie} &
		\rot{\tls} &
		\rot{\redirect} \\
		\cline{1-20}
		\toymail &\Checkmark&\Checkmark&&&&&&\Checkmark&&&&\Checkmark&\Checkmark&\Checkmark&\Checkmark&\Checkmark&\Checkmark&& \\
		\wiggy  &&\Checkmark&&\Checkmark&\Checkmark&&\Checkmark&&&\Checkmark&&&&&\Checkmark&&\Checkmark&& \\
		\barbie  &&\Checkmark&&&\Checkmark&\Checkmark&&&&\Checkmark&&&&&&&\Checkmark&& \\
		\sphero  &&&&\Checkmark&&&\Checkmark&\Checkmark&&&&&&&&&&& \\
		\cozmo    &&&&&&&\Checkmark&&&&&&&&&&&& \\
		\ozobot  &&&&&&&&&&&&&&&&&&& \\
		\osmo  &&&&&&&&&&&&&&&&&&& \\
		\monkey  &&&&&&&\Checkmark&&&&&&&\Checkmark&&&&& \\
		\vtechtab  &\Checkmark&&&&&&&&&&&&&\Checkmark&&&&& \\
		\chip &\Checkmark&&&\Checkmark&&&\Checkmark&&&&&&&&&&&& \\
		\cloudpets &&\Checkmark&&&&&\Checkmark&&&\Checkmark&&&\Checkmark&\Checkmark&\Checkmark&\Checkmark&\Checkmark&& \\
		\cline{1-20}
	\end{tabular}
	\end{adjustbox}
\end{table*}   

In terms of PII collection and contacting third-party servers, all toys we examined collected PII, and 9 out of 11 communicated with one to four ads and analytics servers; see Table~\ref{tab:ads}. Alarmingly, the toys for younger children collected PII more aggressively and frequently, including intrusive PII, such as cross-service and cross-platform unique identifiers and user and devices fingerprinting information; these toys also in general protected PII more poorly. One such toy in our samples (Cloudpets) also displays ad in its companion app. In contrast, toys for older children collected less PII, generally did not require account creation, and had more thoughtful security measures in place. 

\begin{table*}[!htb]
	\small
	\centering
	\caption{Ads and analytics services contacted by smart toys and companion apps}
	\label{tab:ads}
	\begin{tabular}{lccccccccccccc}
		~  &
		\rot{ tangible-analytics.appspot.com } &
		\rot{ googleads.g.doubleclick.net } &
		\rot{ googlesyndication.com } &
		\rot{ googleadservices.com} &
		\rot{ google-analytics.com} &			
		\rot{ data.flurry.com} &
		\rot{ e.crashlytics.com} &
		\rot{ unity3d.com} &
		\rot{ hockeyapp.net } &
		\rot{ api.branch.io} &			
		\rot{ api.segment.io } &
		\rot{ wzrkt.com } &           
		\rot{ads.mopub.com } \\
		\cline{1-14}\\
		\toymail &&&&&&&\Checkmark&&&\Checkmark&\Checkmark &\Checkmark& \\
		\wiggy  &&&&&\Checkmark&&&\Checkmark&&&&& \\
        \barbie  &&&&&&&&&& \\
		\sphero  &&&&&&\Checkmark&&\Checkmark&&&&& \\
		\cozmo    &&&&&&&&&\Checkmark&&&& \\
		\ozobot  &&&&&&&&\Checkmark&&&&& \\
		\osmo  &\Checkmark&&&&&&\Checkmark&&&&&&\\
		\monkey  &&&&&&&\Checkmark&&&&&& \\
		\vtechtab  &&&&&&&&&& \\
		\chip &&&&&&\Checkmark&\Checkmark&&&&&&\\
		\cloudpets &&\Checkmark&\Checkmark&\Checkmark&&&&&&&&&\Checkmark\\
		\cline{1-14}
	\end{tabular}
\end{table*}

Table~\ref{tab:ids} shows the persistent user IDs collected by the toys and companion apps. Apparently, some IDs captured in the transmissions were obfuscated, or their purpose was unclear. In such cases, we assume that if they appear across multiple sessions, like an app ID, then they are deemed to track the user across multiple sessions of the same app. All the Android apps we tested (\toymail, \wiggy, \barbie, \chip, and \cloudpets) transmit the Google advertising ID, even when ad tracking is disabled in the smartphone. These apps, and those that transmit a hardware ID, are deemed to track users across multiple services on the same device. Finally, apps that use cookies such as DSID and IDE on Android~\cite{googlecookies}, or transmit email addresses or other identifying information about the user to analytics servers, are deemed to track users across multiple platforms. On iOS apps, the smartphone's advertising identifier was not transmitted when the ``Limit ad tracking'' was disabled. 

\begin{table}[!htb]
	\small
	\centering
	\caption{Types of persistent user identifiers collected by smart toys}
	\label{tab:ids}
	\begin{tabular}{lcccccccc}
		&&\rot{ \parbox{2.7cm}{Multi-session tracking} } & &\rot{ \parbox{2.7cm}{Multi-service tracking}} &&
		\rot{\parbox{2.7cm}{Multi-platform tracking}} &&\\ 
		\cline{1-9} 
		\toymail  &&\Checkmark&&\Checkmark&&\Checkmark&& \\
		\wiggy  &&\Checkmark&&\Checkmark&&&& \\			
		\barbie  &&\Checkmark&&\Checkmark&&&& \\	
		\sphero  &&\Checkmark&&\Checkmark&&&& \\			
		\cozmo  &&\Checkmark&&&&&& \\
		\ozobot &&\Checkmark&&\Checkmark&&&&\\			
		\osmo  &&\Checkmark&&&&&&\\
		\monkey  &&\Checkmark&&\Checkmark&&&&\\
		\vtechtab  &&&&&&&&\\
		\chip  &&\Checkmark&&\Checkmark&&&&\\
		\cloudpets  &&\Checkmark&&\Checkmark&&\Checkmark&&\\		
		\cline{1-9}
	\end{tabular}
\end{table}

Table~\ref{tab:pii} shows types of PII collected by first-party toy servers. Note that we exclude any PII stored locally on the toy or smartphone. Multimedia may consist of audio (e.g., voice recordings), photos, or video, which may be recorded by the user as part of toy functionality, or uploaded as profile information. We consider a toy to transmit geolocation only at street-level granularity or below, not above (e.g., city). All the toys we examined transmit the user's public IP address, but in addition one (Toymail) sends the private IP address of the toy on the local network.  Toymail is unique also in that it transmits the user's Facebook ID and the sleep/wake cycle of the child.

\begin{table}[!htb]
    \small
    \centering
    \caption{PII collected }
    \label{tab:pii}
    \begin{tabular}{lllllll}
    ~ & \rot{Parent's full name} & \rot{Child's full name} & \rot{Child's age} & \rot{Email address} & \rot{Multimedia} & \rot{Geolocation} \\
    \hline
    \toymail & \Checkmark & \Checkmark & ~ & \Checkmark & \Checkmark & ~ \\
    \wiggy & ~ & \Checkmark & ~ & \Checkmark & ~ & ~ \\
    \barbie & ~ & ~ & ~ & \Checkmark & \Checkmark & ~ \\
    \sphero & ~ & ~ & \Checkmark & ~ & ~ & ~ \\
    \cozmo & ~ & ~ & ~ & ~ & ~ & ~ \\
    \ozobot & ~ & ~ & ~ & ~ & ~ & ~ \\
    \osmo & ~ & ~ & ~ & ~ & ~ & ~ \\
    \monkey & ~ & ~ & ~ & ~ & ~ & ~ \\
    \vtechtab & \Checkmark & \Checkmark & ~ & ~ & \Checkmark & \Checkmark \\
    \chip & ~ & ~ & ~ & ~ & ~ & ~ \\
    \cloudpets & \Checkmark & \Checkmark & \Checkmark & \Checkmark & \Checkmark & ~ \\
    \hline
    \end{tabular}
\end{table}

\section{Recommendations}\label{sec:recommendations}

In this section, we list our recommendations for makers of connected toys, beyond following existing safety standards and privacy regulations.  

\subhead{Fixing vulnerabilities} Toys, like other more generalized IoT devices, should have a process for addressing vulnerabilities. This includes a way to disclose vulnerabilities, including: a formalized reporting mechanism, a simple link to a form on their web site, or subscription to a bug bounty program, like HackerOne (\url{hackerone.com}) or BugCrowd (\url{bugcrowd.com}). At the very least, vulnerability disclosure should not be discouraged by prohibiting under the terms of use. There should be a short turnaround between identifying vulnerabilities and fixing them in the firmware or app software where they are located. Finally, it is critical to be able to push out new firmware or software updates and patches as soon as they are available. There are existing mechanisms to update companion apps through the corresponding Android or iOS app store, but pushing out firmware updates to the toy is slightly more challenging. Toys with Wi-Fi connectivity should periodically check for updates, and companion apps should check at the launch time for available updates. However, relying on the user to manually launch the app extends the window of potential exploitation between discovery of the vulnerability and time it is patched, so it is preferable for apps to have push notification functionality enabled. For this reason, companion apps should request permission to push notifications on first launch, and clearly state why this can improve the security and safety of the toy. (It is also incumbent on the app to not abuse the permission with unnecessary notifications lest the user disable it.)

\subhead{Limiting data collection, and improving storage and communication security} Toys should limit the collection, storage, and transmission of personal data to only what is strictly necessary. Toys should not store personal information such as voice recordings, photos, videos, or any personal information that identify users/children, internally on the toy's flash memory; if it is necessary for toy functionality, then it should be stored encrypted. Transmission of personal information should always occur over secure communication channels such as TLS.
Toys and companion apps should use TLS for all internet traffic, with strong cipher suites and recent TLS versions (e.g., TLS 1.2), 
patched against known TLS protocol vulnerabilities. Wherever possible, host whitelisting should be used, hard coded in the smart toy firmware and the companion app, as it can mitigate phishing and MitM attacks. Certificate pinning can also help prevent MitM attacks.   

\subhead{Securing Bluetooth and WiFi connections} Toys with wireless capabilities should take particular care to secure pairing and connection. If the toy uses Bluetooth, it should use Bluetooth encryption. Using open Bluetooth can allow unauthorized access to the toy and PII stored on it. Toys should also use dynamic Bluetooth MAC addresses to avoid the possibility of tracking children's locations. WiFi-enabled toys should be provisioned in a secure manner, with WiFi credentials supplied securely over an out-of-band channel.

\subhead{Fine-grained access control} Toy companies that provide links to PII on public servers, such as voice recordings, should also grant parents the ability to select whether the PII is public or private. Even when links are randomly generated, if they are sufficient to access PII without any additional authorization, it still can pose a privacy threat. We also recommend allowing the parent to configure whether such links are public or private, defaulting to private. Finally, these links should expire after a reasonably short period of time.

\subhead{End-of-life} Toys should provide a graceful means of scrubbing identifying and personal data in the event of end-of-life, loss, theft, or transitioning owners. This may mean restricting the amount of personal data stored locally on the toy to only what is strictly necessary, and requiring some form of authentication to access whatever data remains. 
Toys that require the creation of online accounts tend to collect more personal information than toys that do not, at least in our sample, but also provide a convenient interface for letting a parent log in through a browser and dissociate the toy from the account if necessary. This ensures that an unauthorized third party with access to the toy cannot access the related account and any personal data stored within it, or use it to communicate with the child.

\subhead{Advertisement and tracking} We strongly caution against ad display in companion apps, especially as these apps form part of a purchased toy and thus should not require support through ad revenue. In addition, ads and analytics services should be used sparingly with children~\cite{austin1999}, and when used, should not collect any PII. Various other measures can be taken to protect the user's identity when using analytics services; Google Analytics, for instance, provides a means of anonymizing IP addresses~\cite{googleipanonymization}. Flurry Analytics prohibits the use of its services with children under the age of 13 (in accordance with COPPA) unless its restricted feature set limiting user profiling is used~\cite{flurryrestrictedset}.

Apps should respect users' request to disable ad tracking. This is an issue particularly on Android platforms.
Specifically, we observed multiple instances of Android apps flouting the user's preferred ad tracking setting, and sending the ad identifier regardless of whether it was disabled. By contrast, iOS app developers cannot access a valid ad identifier on a device that has ad tracking disabled~\cite{appleadid}. 
In the Android model, a single system call returns both the ad identifier and the limit ad tracking setting, and the developer is assumed to not abuse them~\cite{googleadid}. 

\section{Conclusions}\label{sec:conclusions}
In this paper, we examined the security and privacy practices of eleven popular smart toys on the market. Our findings show widespread use of data collection by toy makers and third party analytics servers, with almost all toys embedding analytics services within their apps, and many of them transmitting advertising identifiers even when ad tracking is disabled by the user. Several toys do not take adequate measures to protect sensitive data, and some toys do not protect locally stored PII at all, making it trivially accessible to anyone with physical access. Children's physical safety may also be at risk, as all Bluetooth-capable toys we examine advertise a static MAC address, making it possible to track them through physical space. Finally, several toys are vulnerable to exploits that can result in handing an attacker control of the toy or some of its functionality, potentially causing harm to a child.  

\section*{Acknowledgments}
This work is supported by a grant from the Office of the Privacy Commissioner of Canada (OPC) Contributions Program. 
\bibliographystyle{abbrv}

\appendices

\section{Results Analysis of Selected Smart Toys} \label{sec:smarttoys}
In this appendix, we present individual case studies for the toys mentioned in the main body of the paper. 

\subsection{\toymail}
\begin{figure}[h!]
	\centering
	\includegraphics[scale=0.32]{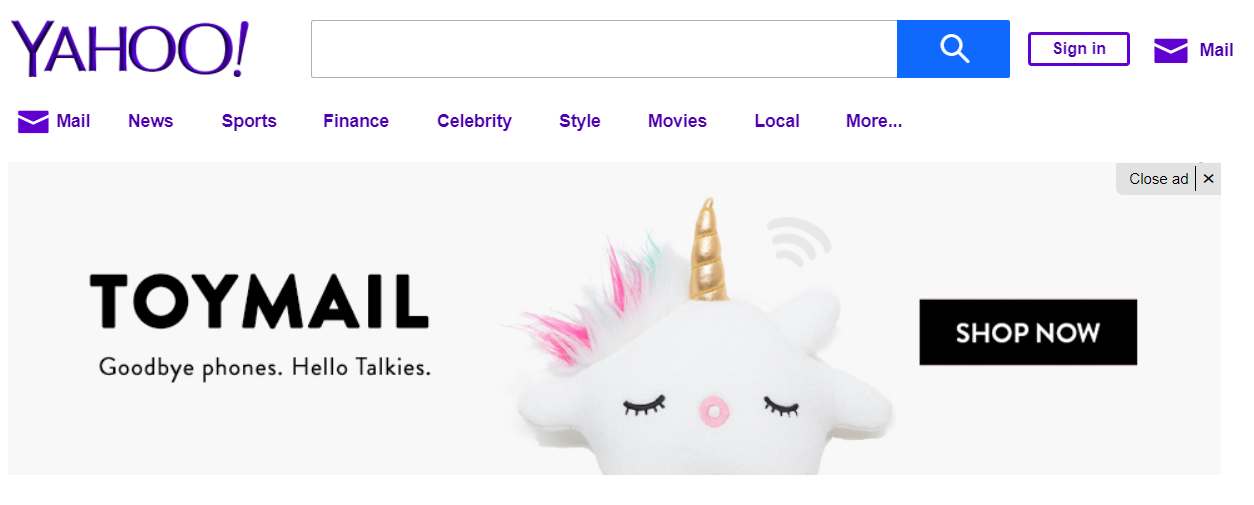}
	\caption{Tracking example: Toymail tracking users across different platforms - \toymail\ ad shown in a PC while the user is browsing Yahoo!}
	\label{fig:toymail}
\end{figure}

\toymail\ allows children to exchange voice messages with parents, relatives and friends using the companion app or another \toymail. Parents can configure the toy with an approved list of contacts. Voice alerts notify children when they have new messages, and they can click the play button to listen. 
\subhead{\pii}
The companion app sends PII to app.toy mailco.com including login credentials; email address; parent profile ID; child name, date of birth, and photo; friend profile ID, name, and profile picture; and voice messages.
The app communicates with multiple ads and analytics services which collect personal information as follows. E.crashlytics.com collects smartphone device information and app crash information, along with smartphone hardware ID, Google ads ID, and app installation ID. Api.branch.io collects smartphone device and OS information, Toymail user ID, smartphone hardware ID, smartphone fingerprint ID, IP address, identity ID, and email address. Api.branch.co verifies whether the smartphone is a real device or an emulator. Wzrkt.com sets a unique ID for the user which it collects while the app is running, allowing tracking the user across different sessions. It also collects the Google Ad ID (allowing tracking the user across different services), smartphone device information, and telephone service carrier information. Api.segment.io collects \toymail\ user ID, name, email address,  device hardware ID, Google ads ID, smartphone device and OS information, data network device status (WiFi, Bluetooth, cellular), and phone carrier name.
None of these ads and analytics servers respects the device-wide ad tracking setting, and all collect PII regardless of whether it was disabled. Unsurprisingly, we found that using the toy resulted in targeted ads even on different platforms by setting unique identifiers (e.g., identity ID) and correlating them with other personal information (e.g., email address), as shown in Fig.~\ref{fig:toymail} which depicts a \toymail\ ad in a PC hosting Windows, on a different WiFi network.

Once the toy connects to the WiFi network, it communicates with imp07a. boxen.electricimp.com to receive a list of allowed contacts. When the child selects a contact and presses the message button, the toy begins recording. When the child presses the message button again, the toy transmits the voice recording to the server, which stores them in Amazon AWS storage.
The toy keeps up to ten voice messages locally which are received from parents and contacts. It also stores \toymail\ child ID, toy ID, contacts' names, and contacts' \toymail\ IDs, in addition to WiFi SSID and password. 

\subhead{\secmes}
The app connects to all servers over TLS. The app.toy mailco.com server certificate is signed by Go Daddy and uses a 2048-bit RSA key. The app maintains its own list of strong cipher suites, and uses certificate pinning to prevent MitM attacks. It stores hashes of the certificate chain public keys to bind the server certificate with the original issuer. 
However, the toy suffers from a number of \emph{\tls}. The imp07a.boxen.electricimp.com server certificate is signed by the custom root CA impca, and both certificates are sent to the toy. The server certificate is valid for 20 years and it uses a 1024-bit RSA key whose certificate common name is different from the server's domain name. The toy uses a client certificate to authenticate to the server. It is signed by impca, uses a 2048-bit RSA key, and is valid for 100 years. The app communicates with two analytics and ads servers vulnerable to RC4 and POODLE attacks, with support for SSL 3.0 and weak cipher suites, exposing \toymail\ to \emph{\third}. Neither the session cookie expiration date nor secure flag are set, leaving the toy vulnerable to \emph{\cookie}. 

\toymail\ is rated to offer \emph{\parental}. When a child sends recordings to contacts other than the parent, they are stored in the contact's account, leaving the parent without means to review or delete them. Even when a parent blocks a contact from the child's contact list, that contact can still access the child's personal information. The server sends the child's personal information to the contact's app accompanied by a ``blocked'' flag (equal to ``-1'') indicating that the recipient has been blocked. The app removes the entry corresponding to the child from the app interface so they can no longer exchange voice messages; however, the contact still receives the child's personal information each time the app is launched including: name, date of birth, photo, voice recording of the name, parent \toymail\ ID, toy ID, wake up time, sleep time, and ``isOnline'' flag showing the state of the toy. 

With physical access to the toy, anyone can access these messages and even add contacts to the contact list without authentication, making it vulnerable to \emph{\local} and \emph{\configphys}, respectively. In fact, there is no mechanism to delete locally stored voice recordings other than configuring the toy to use another account. 

\toymail\ is designed to be \emph{\remains}. Switching off requires removing the batteries, accessible only by removing the back cover using a screwdriver. Given that the toy includes sensitive devices like a microphone and is connected to the Internet, we deem this a significant potential threat to children's privacy and security.

\subsection{\wiggy}
\wiggy\ is a Bluetooth-connected toy that allows parents to set tasks, create goals, and send rewards to their children through a companion app. Children are notified they have new tasks either by the toy, which gives alerts and states the tasks, or by checking their accounts through the app. Children confirm through the app that they have completed a task, and upon parental approval virtual funds are transfered to their account. A child can set goals (e.g., buying a mobile phone for \$100). When the balance of the child's virtual account achieves the target value of the goal, the child may redeem it from the parent.
\subhead{\pii}
The app communicates with the server api.kii.com, sending parent login credentials (email address and password or session cookie), user ID, children's profile pictures, tasks set by parent, tasks completed by children, and children's goals.  
The app also communicates with ads and analytics services. Unity3d.com collects an app ID, device unique ID, user ID, session ID, and device and OS information. Moreover, even when ad tracking is disabled, unity3d.com still collects an advertising ID uniquely identifying the user across different services, and app usage information including app start time and duration.  
The app also connects to google-analytics.com, which collects analytics data and applies obfuscation techniques before sending them. 

\subhead{\secmes}
The app's connection to api.kii.com appears to use TLS in a secure way. The app relies on the OS trusted store to check the server certificate. The server uses a certificate signed by well-known certificate authority (Go Daddy Secure Certificate Authority - G2) and signed by a 2048-bit RSA key using SHA256. The server public key uses a 2048-bit RSA key and the certificate is valid for a limited period (expiring in October 2019). The server has six cipher suites that support forward secrecy and they are at the top of the server's cipher suites list. The server is patched against vulnerabilities including POODLE and Heartbleed, Ticketbleed, and does not support TLS compression, which causes vulnerabilities like CRIME. 
Although the app uses HTTPS to communicate with the server api.kii.com, the server does not force using HTTPS in the case of flipping HTTPS to HTTP, leaving all communication unencrypted, and making the toy vulnerable to \emph{\comm}. We use the Burp Suite repeater tool to flip all app requests to the server on the fly to HTTP instead of HTTPS, and as a result, the app connects to the HTTP version of the server and receives PII in plaintext.

A determined adversary can take over a parent's account in several ways. The first scenario is a result of \emph{\cookie}. All session cookies expire January 19, 2038 - a long period of time during which any adversary who steals the cookies may use them to access the parental account. In a second scenario, an \emph{\bruteforce} vulnerability allows an attacker to gain access to the parent's account. This was verified using the Burp Suite intruder tool to brute force accounts owned by us. 
However, the most serious vulnerability we uncovered, \emph{\confignearby}, enables an adversary to access PII or assign tasks to a child simply by being within Bluetooth coverage. Anyone within Bluetooth range can install the companion app, pair with the toy, and issue tasks, both new and predefined. These new tasks would appear on the child's account just as if they had been assigned by the child's parents. The toy announces predefined tasks in what to the child is a familiar and trusted voice, increasing the likelihood that the child will fall prey to the attack. The attacker may also add a valuable (but fake) reward to encourage the child to perform the task. An adversary needs to take over the parent's account (for instance by brute forcing the parent's password) to be able to assign new tasks to the child, but does not need to do so to assign predefined tasks, including some that encourage the child to leave the house like ``pull the weeds'', ``rake the leaves'', ``take out the trash'', and ``walk the dog''. The toy also broadcasts a static Bluetooth MAC address, making it vulnerable to \emph{\bluetooth}.

\subsection{\barbie}
\barbie\ is a smart doll designed to conduct interactive conversations with a child. The toy connects to the Internet directly to send the child's voice recordings to the toy server, which applies voice recognition techniques to respond to the child. The toy has a companion app that a parent can use to configure the toy.
\subhead{\pii} 
The companion app transmits the following information to api.2.toytalk.com: the smartphone manufacturer, device name and install ID (unique per device), account ID (unique per user), app version, user consent flag, OS name, and OS version (see also Table~\ref{tab:ads}). The app also sends login information including email address and password, as well as the child's date of birth and any other important dates the parent has specified. 
We did not observe traffic to any ads or analytics services. The toy does not store the child's voice recordings, but streams them directly to the cloud. The toy stores the WiFi network SSID and passphrase, as well as the parent account ID that is used to access the parent profile and listen to recordings. 
On the other hand, the app stores sensitive information in plaintext, including a session cookie authenticating the user to the toy server, the parent email address, and profile ID. 

\subhead{\secmes}
The app does not rely on the OS CA store to verify the server certificate. The custom CA store, which is included in the app's assets directory, contains one certificate for ``Toytalk ca'', a self-signed certificate used to sign the server certificate. In addition, the app uses a Bouncy Castle PKCS\#12 certificate to authenticate to the server.
The app connects only to servers in its whitelist; at the time of testing, these were test.2.toytalk.com and api.2.toytalk.com, although we did not observe any communication with the first server, which seems to be for testing purposes. 
The app communicates with api.2.toytalk.com over TLS using the OS TLS implementation, which may contain weak cipher suites or vulnerable practices specifically in older OSes such as Android 4.4. 

There are several scenarios in which an adversary can exploit \emph{\tls} to steal the parent's credentials and access the child's recordings. The client, server, and self-signed root certificates use RSA with 1024-bit keys which are being phased out~\cite{Mozilla2014}. The server certificate private key can be used to decrypt the negotiated session key, and the root CA key can be used to sign certificates as api.2.toytalk.com to steal the parent's credentials.  

The toy communicates with three servers: firmware.toytalk.com, puppeteer.toytalk.com, and storage.toytalk.com. When \barbie\ is switched on, it connects to the WiFi network and initiates a connection to firmware.toytalk.com over TLS. While the child presses and holds the ``talk'' button, \barbie\ streams the child's speech over a TLS-protected channel to puppeteer.toytalk.com. Once the ``talk'' button is released, puppeteer.toytalk.com sends the link to the new recording to storage.toytalk.com, where it is stored. Analysis of the three servers shows that two of them, namely firmware.toytalk.com and storage.toytalk.com, support weak cipher suites (TLS\_RSA\_WITH\_3DES\_EDE\_CBC\_SHA with TLS 1.0). Bhargavan et al.~\cite{weak2016blockciphers} found that ciphers using 64-bit blocks (e.g., 3DES) are vulnerable to secret key disclosure.

The official website toytalk.com can be used by parents to review the child's recordings. The server certificate uses an RSA 2048-bit key and is signed by Amazon using SHA256 and an RSA 2048-bit key. The server does not support weak cipher suites and uses cryptographic libraries patched against known cryptographic attacks. However, there is no restriction on the number of login attempts, and the login can be bruteforced. We exploited this \emph{\bruteforce} vulnerability during the course of testing to bruteforce our own account. This flaw was originally reported in January 2016\cite{Somerset2016barbie} and our experiments show that it is still not fixed.

To investigate communication between the toy and the app, we use the following specific setup for \barbie. We use two WiFi adapters on our test machine, one in AP mode as an impostor Hello Barbie, and the second to connect to the real Hello Barbie, exploiting its unprotected hotspot. We route all traffic from the impostor to the real toy. In the app, both toys, real and impostor, are listed, and we connect to the impostor. The smartphone hosting the app, meanwhile, is configured to forward all traffic through a Burp Suite proxy. This configuration requires the app to be patched to accept certificates signed by Burp Suite CA, and the app client certificate bundle added to Burp Suite as the client certificate. This allows the app to authenticate itself to the toy, enabling us to intercept communication between them. The toy authenticates to the app using a certificate issued to 192.168.10.1, signed by ``ToyTalk CA'', and valid to 2030. Once the TLS handshake is complete, the app sends the parent account ID and WiFi configuration, including WiFi SSID and password, to the toy.

The toy can be configured using any companion app, and no parent authentication is required to pair with the toy. Because the toy uses the parent's account ID as the sole way to relate the child's profile, including voice recordings, to the parent, an adversary need only reconfigure the toy to use the adversary's account to access all subsequent recordings made by the child. 

There are several scenarios in which the toy is vulnerable to \emph{\confignearby}. In the first scenario, 
an adversary who knows the WiFi credentials can configure the toy with their account ID; in the others, the adversary does not require the WiFi credentials. While the toy is in pairing mode, it broadcasts an open network with SSID ``Barbie-950'', making it temporarily vulnerable to \emph{\hotspot}. An unencrypted hotspot could allow an adversary to conduct ARP spoofing, assuming a MitM position between the app and the toy. In this scenario, the legitimate toy and the app will unwittingly treat the adversary as the destination for all traffic, allowing the adversary to sniff the parent account ID and the WiFi credentials. We deem this a minor risk as the vulnerability is only present for a limited duration (during pairing), and an attacker must be within close proximity. However, the toy can be accidentally put in pairing mode during the course of normal play, as it requires the child to simultaneously press two easily accessible buttons on the toy. Once it is in pairing mode, an adversary within wireless range can download the app, configure the toy to connect to a different WiFi network under their control, and then configure it to use the adversary's account. 

\subsection{\sphero}
\sphero\ is a Star Wars character-branded spherical robot that can be remotely piloted via a companion app. The \sphero\ is one of a line of spherical robots designed for both recreational and educational use. It can be controlled programmatically using the beginner-friendly Sphero SPRK Lightning Lab smartphone app, or a variety of both official and unofficial SDKs for multiple platforms. The toy features an on-board gyroscope and accelerometer, and is designed to be controlled by compatible BLE-enabled devices. Although \sphero\ has been designed to be controlled in a variety of ways, including programmatically, our study focused primarily on the most common (and, to a child, intuitive) scenario: running the companion \sphero\ app on a smartphone. 

\subhead{\pii}\label{subsec:sphero_pii} 
The \sphero\ companion app does not require the user to create an account, and does not require a username or email address to operate. 
The app prompts the user to enter their age, but does not solicit any other personal information, and transmits a flag indicating whether the user is under the age of 13, to both a Sphero-hosted server at gosphero.com and to Flurry Analytics. The age 13 is significant as it is the cut-off age for COPPA rules. However, we did not find that indicating an age under 13 had any other impact on either the frequency or content of communications between the app and servers hosted either by Sphero or third-parties. Regardless of whether the user was identified as being under 13, the app transmits a unique app identifier and the MAC address of the toy. Other than analytics data, the app communicates primarily with sphero.com and gosphero.com. The app queries sphero.com for firmware updates and media files for use within the app, some of which are hosted on Amazon AWS. It sends detailed toy usage data to gosphero.com, including a unique app ID that remains constant across multiple deletions and re-installations of the software; the toy's Bluetooth MAC address; and timestamped events related to toy speed and collisions. The timing of these transmissions suggests they are event-based, triggered by, for instance, toy collisions.

The \sphero\ app sends analytics data to Flurry Analytics (formerly Data Flurry), a subsidiary of Yahoo!. While toy-specific data, such as speed and collision data, is sent to gosphero.com, app-specific data, such as buttons pressed, commands accessed, and slider values, are sent to Flurry Analytics.
Each HTTP post to Flurry Analytics includes a device-wide Flurry ID, a unique app ID, and the MAC address of the toy. The Flurry ID is unique to the device, even across unrelated apps by different developers, all using Flurry Analytics. By default, each call to Flurry Analytics sends richly detailed device information, including: processor; OS; memory, disk, and battery usage; and geographical location such as time zone, locale, and city. 
Flurry Analytics' ToS prohibits developers from sending PII, such as a user's ID and email address, unless it is cryptographically obscured using a one-way hashing function~\cite{DataFlurryToS}. However, the granularity of the data captured by Flurry Analytics can be used to fingerprint the device and track user behavior across different apps on the same device that also use Flurry Analytics' services. Flurry Analytics' web site states that it uses this data to build a demographic profile of the app user, grouping users into such niche demographics as ``Business Travelers, Pet Owners, and New Moms, among many others''~\cite{DataFlurryDemographics}.
At launch, the \sphero\ app sends device hardware information to app gaming engine Unity 3D (unity3d.com). 

\subhead{\secmes}
The \sphero\ can only be controlled via BLE. The companion app only allows Bluetooth pairing with \sphero\ when the device running the app is within close proximity to \sphero. Simply being within BLE range is insufficient to pair with the toy when using the companion app; in general, proximity pairing requires the Received Signal Strength Indicator (RSSI), a measure of the strength of the signal between device and toy, to be within a circumscribed range. We rate this practice as tending to increase the security of the toy overall; 
however, the \sphero\ is subject to the \emph{\bluetooth} vulnerability. It broadcasts a static Bluetooth MAC address, which appears to remain fixed over the lifespan of the toy, even across power cycles - making the toy, and hence a child using it, trackable. The toy advertises its MAC address over BLE even when placed on its charging cradle and not in use, maximizing the risk of unwanted detection. This flaw is exacerbated by the \emph{\remains} vulnerability.

Only a single companion app can connect to a single \sphero\ at a time. However, any PC or laptop running the \sphero\ SDK can connect to the toy and control it if it is within BLE range, regardless if the toy is currently maintaining an active connection with another device, a \emph{\play} vulnerability. The only information required to take control of a \sphero\ using this method is its Bluetooth MAC address, which is trivial to obtain when the device is within Bluetooth range (given that it is constantly advertising its presence). 
In our test scenario, we used a node.js SDK to remotely pilot the BB-8 and make it behave in an erratic fashion, rolling wildly in random directions and changing its on-board LED color unpredictably. Listing~\ref{lst:sphero_node.js} shows a snippet of node.js code used to control BB-8 from an unauthorized PC. Our intent was to simulate a scenario where a malicious actor outside visual range could cause a child's toy to behave uncontrollably, thereby causing the child distress. Such an actor would only have to be within a 33-metre range, the theoretical maximum range for BLE. We do not rank this threat as either particularly likely or particularly damaging, and it does not leak any privacy-related information about the child to an attacker. However, we can envision a scenario in which an attacker in a public space uses this method to navigate the toy in the attacker's direction, prompting the child to follow the toy and be lured towards the attacker.

\begin{lstlisting}[float,floatplacement=H, language=javascript, caption = Snippet of node.js code to control BB-8, label = lst:sphero_node.js]
bb8.connect(function() {
// roll BB-8 in a random direction, changing direction every second
setInterval(function() {
var direction = Math.floor(Math.random() * 360);
bb8.roll(150,direction);
// set a random color
bb.randomColor();
}, 1000);
});
\end{lstlisting}

\subsection{\cozmo}
\cozmo is an intelligent programmable bot with onboard speaker, camera, and inbuilt facial recognition. Cozmo comes with 3 programmable, LED-colored cubes. \cozmo\ is interactive, and can be played with using a companion app, or programmed via a Python-based SDK, available on GitHub.
\subhead{\pii}
When the \cozmo\ app is launched, it transmits unique app and session identifiers to HockeyApp.net, an analytics subsidiary of Microsoft, along with device fingerprinting information such as the device model, locale, OS version, and screen resolution. It also sends obfuscated data to a server hosted on Amazon AWS. \cozmo\ does not require account creation to operate. 

\subhead{\secmes}
\cozmo\ engages in \emph{\tls}. The server hosted on Amazon AWS supports weak cipher suites, including 3-DES, and 1024-bit Diffie-Hellman key exchange mechanisms, weakening perfect forward secrecy.
Because \cozmo\ can transmit a live video stream via its onboard camera, we investigated what measures were taken to harden it against attacks. A client device running the companion app must connect to Cozmo's wireless access point over 802.11g to establish a play session. Cozmo's SSID is broadcast only for a limited time when placed in its cradle, after which it is no longer available to connect. The network is secured over WPA2 Personal using a 12-character alphanumeric shared passphrase displayed on Cozmo's LCD screen during the connection period. The passphrase is fixed for each Cozmo, does not vary over time, and cannot be changed, making this practice slightly less secure than it could be otherwise.
The companion app host device does not have Internet connectivity during a play session as it is connected to Cozmo's hotspot, so even a compromised device running the Cozmo companion app has a minimal risk of re-transmitting a live video feed to an attacker's web site unless it is somehow stored locally for re-broadcast later, a scenario we regard as unlikely.
Cozmo may be programmed from a desktop or laptop computer using the Python SDK under the following conditions: the computer must connect via USB to the smartphone or tablet running the Cozmo companion app, and the companion app must be explicitly configured to run in SDK mode. In this configuration, only the smartphone and not the computer connects to Cozmo's access point. Furthermore, only a single companion app can connect to Cozmo at any given time. 
These countermeasures combine to minimize the risk of an attacker surreptitiously intercepting Cozmo's video feed, even if they are in possession of Cozmo's SSID and password. 

During an active play session, Cozmo connects to the companion LED cubes over BLE. After the play session, neither Cozmo nor the cubes continue to advertise over BLE, minimizing the associated privacy risks. However, the Bluetooth MAC addresses of Cozmo and the LED cubes are static, increasing the risk they can be used to track a child.

\subsection{\ozobot}
\ozobot\ is a small bot with motors and an optical sensor that detects and responds to changes in color of the surface it happens to be on. The child ``programs'' the Ozobot by drawing different colors on paper or on the companion app, and the Ozobot spins, speeds up or down, and changes direction accordingly.

\subhead{\pii}
\ozobot\ does not require account creation to operate, nor does the companion app collect any personal information beyond what can be acquired during normal app use. The Ozobot companion app communicates primarily with analytics servers, and all connections it initiates are over TLS. Ozobot posts analytics data to unity3d, the app game engine platform, sending app, user, session, and device identifiers; platform type and version; the app-related event that triggered the post, such as ``appStart'', ``appRunning'', ``appStop''; and a timestamp and local time offset. There are additional posts to app-measurement.com, registered to Google DNS Admin, and googleapis.com containing app and device identifiers.

\subhead{\secmes}
As no PII is collected beyond device identifiers and analytics data, and there are no sensitive sensors on the device, applicable security measures here consist of proper TLS use. All servers contacted by the app conform to proper TLS security practices.

\subsection{\osmo}
\osmo\ is modular, consisting of a suite of toys, each sold separately with a companion app developed for use exclusively with Apple iPads. All toys in the Osmo suite use a mirrored attachment that hangs over the front iPad camera in portrait orientation, reflecting the surface in front of the iPad as it sits upright on a base. The child's interaction with the Osmo toy on this surface is captured by the camera and processed by the companion app in a module-specific way. We performed tests on Osmo Tangram, Numbers, and Letters.
\subhead{\pii}
Osmo toys do not themselves have any sensors, but require the companion app to have permission to access the camera. Because the app is used with the Osmo reflector, during active use the front camera transmits only the surface in front of the iPad.
Like \ozobot, \osmo\ does not require account creation to operate, nor does the companion app collect any personal information beyond what can be acquired during normal app use. All connections initiated by the \osmo\ companion app are over TLS. The \osmo\ app connects to playosmo.com, sending a persistent identifier and retrieving dynamic localizations displayed in the app interface. It sends analytics data to tangible-analytics.appspot.com using the same persistent identifier sent to playosmo.com, as well as obfuscated session state data.

\subhead{\secmes}
As no PII is collected or sent to Osmo-hosted servers beyond device identifier and other analytics data, and \osmo\ does not contain any sensitive sensors, applicable security measures here consist of proper TLS use. Both playosmo.com and tangible-analytics.appspot.com engage in secure TLS practices, with proper certificates and servers patched against common attacks.

\subsection{\monkey}
\monkey\ is one of a related line of Smart Toys by Fisher Price. The toy is interactive, featuring a microphone, speaker, and voice recognition, and an accelerometer that detects if it is being thrown in the air. The toy has limited image recognition, mainly for use with accompanying activity cards that can be used in lieu of the companion app.

\subhead{\pii}
On initial app launch, the parent is prompted to create an account. The app scans for available WiFi networks and asks the parent to select one and input its passphrase. It generates an encrypted, base64-encoded QR code to display to the toy, which scans it using image recognition and connects to the WiFi access point.
Both the toy and the companion app communicate with Crashlytics and smarttoy.org. The app transmits to Crashlytics its status in response to app events such as terminate, enter foreground, enter background, is inactive, is active, along with device and session identifiers.

The app sends and retrieves extensive data from smarttoy.org, sometimes megabytes of data in a single session. On closer examination, we find that multiple times in a single session the app retrieves a dictionary of hundreds of words and phrases localized to English, German, Spanish, Italian, and French. Some of these are misspelled, grammatically incorrect, inappropriate, and/or profane; examples include:                    ``because their [sic] all dead'',
``why are you single'',            
``loading a gun'',
``I was a victim''
``his daddy was a mummy'',           
``fuck you'',
``motherfucker'',
``bitch''.
The toy does not respond to these phrases when vocalized, and we could not determine their purpose, but each instance of the dictionary is 3 MB of ASCII text, and we observed 10 of these transmissions in a single session.
The toy contacts pro.ip-api.com, a geolocation service, and sends device and user ID to 42.121.18.150, a server registered to Caoyi in China.

\subhead{\secmes}
While the \monkey\ establishes its own WiFi connection to a wireless access point, the connection between the app and the Monkey is over Bluetooth. The Monkey broadcasts a static MAC address and local name ``STParent'', making it susceptible to the \emph{\bluetooth} vulnerability. The toy is susceptible to the \emph{\comm} vulnerability. Some web traffic is sent unencrypted over HTTP, including a request to the IP address 42.121.18.150 located in China with user and device ID in the query string. The user agent string from this HTTP request reveals that the toy is running Android 4.4.2, build KVT49L.
\subsection{\vtechtab} 
\vtechtab\ is a WiFi-enabled tablet targeted to children. It is based on Android 4.4.2 and features out-of-the-box child-friendly apps and whitelisted websites. Websites can be added to the whitelist only by an authorized parent account. The tablet has a camera and microphone. As a standalone tablet, the \vtechtab\ is the only toy we examined that does not have a companion app. We installed our MitM proxy's certificate through the parent account interface and were thus able to intercept TLS communication between the toy and remote servers.
\subhead{\pii}
The \vtechtab\ does not transmit any PII to either VTech servers or ads and analytics servers during normal, non-web browsing use. As an Android-based tablet, it persists to local storage data generated during each user session, such as logs, images, notes, and videos. 

\begin{figure}[!ht]
	\centering
	\includegraphics[scale=0.25]{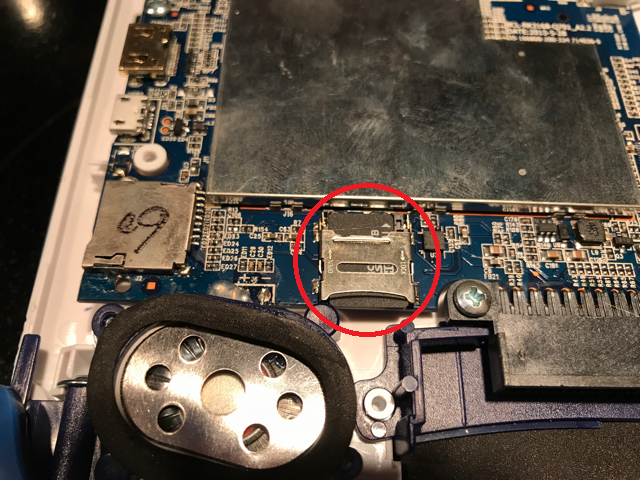}
	\caption{\vtechtab\ internal SD card}
	\label{fig:vtech_sd}
\end{figure}

\subhead{\secmes}
The \vtechtab\ suffers from a number of \emph{\local} vulnerabilities. Children's profiles on the tablet are not password protected. In fact, the launch screen displays the usernames and avatars of all the profiles configured on the device, and clicking the associated avatar logs in as that user, without any authentication required. Once logged in, it is trivial to access and modify all saved PII, including photos, videos, audio recordings, and notes. However, with physical access to the device, it is unnecessary even to log in. The device can be booted in recovery mode and a root ADB shell obtained. By default, a limited number of shell commands are available, but we use the ADB utility to install BusyBox, an executable available in APK format that combines pared down versions of many common *nix utilities. Once BusyBox is installed, we exfiltrate all personal data on the device, including logs, photos, and videos for all profiles configured on the device. Using the same mechanism, a malicious person can alter or add photos and videos that include content that is inappropriate for children. 
Additionally, using the root ADB shell, we have copied the firmware image, modified it, and successfully rebooted using the altered image. An attacker with physical access to the device can alter the image in such a way as to obtain a reverse TCP shell, with the ability to turn on the camera at will.
Finally, the vulnerability noted by Pen Test Partners~\cite{ptpvtech2015extraction} is still present. All profile data is stored, in unencrypted form, on a mounted internal SD card (distinguished from the external SD card), shown in Fig.~\ref{fig:vtech_sd}, and can simply be removed.

The parent account on the tablet is protected by a simple 4-digit PIN that can be easily guessed, as there is no limit to the number of login attempts, making it susceptible to the \emph{\bruteforce} vulnerability. We rank the severity of this vulnerability as minor, as it requires physical access to the tablet.
Moreover, the \vtechtab\ suffers from the \emph{\comm} vulnerability. While much of the communication between the \vtechtab\ and remote servers is encrypted over TLS, some of it is not. Firmware updates are fetched over HTTP, allowing a MitM attacker to downgrade the firmware image. Alarmingly, many interface elements, such as icons and other non-PII media displayed as part of the user interface, are transmitted over HTTP, making them susceptible to a MitM attack and replaced with inappropriate content.

\subsection{\chip}
\chip\ is a robot dog that responds to voice commands and touch. \chip\ comes with a Smart Ball and Smart Bed that doubles as a charging dock, and responds to commands from a companion watch called a Smart Band or the \chip mobile app.
\subhead{\pii} 
The \chip\ companion app sends analytics and bug-related data to Flurry Analytics and Crashlytics. Flurry Analytics collects a hardware unique ID and ad ID, as well as toy usage information, smartphone and OS information, telephone carrier information, approximate location, and smartphone state information including battery remaining percentage, battery charging  state, memory available, external and internal disk sizes and available spaces. Crashlytics collects ad ID, app installation ID, and smartphone device information, in addition to other obfuscated data.

\subhead{\secmes}
\chip\ adopts strong TLS practices. The app does not use TLS 1.0, and the servers are patched against POODLE, Downgrade, and Heartbleed attacks. \chip\ is not vulnerable to CRIME since data compression is not supported in the app or the servers. The app uses its own list of cipher suites which does not contain weak ciphers or short keys, and it supports only TLS version 1.2. All cipher suites support forward secrecy. Flurry Analytics and Crashlytics certificates use RSA 2048-bit, and are issued by well-known issuers with secure signature algorithms. Certificates validity periods are limited, and they are not revoked. Although both servers support TLS 1.0 and a weak cipher suite (TLS\_RSA\_WITH\_3DES\_EDE\_CBC\_SHA\-112), since they are not supported by the app, we consider the app not to suffer from \emph{\tls}.

\begin{figure}
	\centering
	\includegraphics[scale=0.34]{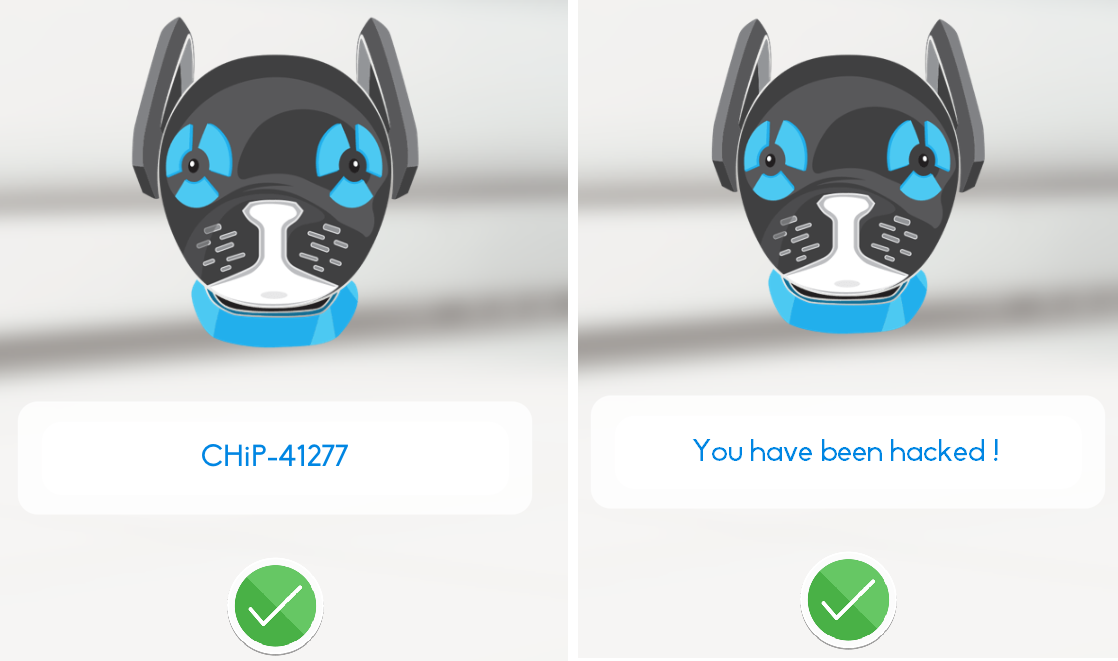}
	\caption{Maliciously changing \chip's name}
	\label{fig:chip_change_name}
\end{figure}

On the other hand, \chip\ is vulnerable to \emph{\local}. All components of the toy can be paired with the RamBLE utility without authentication, allowing us to access and modify toy information that includes manufacturer name, model number, serial number, firmware revision and battery level. An adversary could also modify the name of the toy to something inappropriate, which would appear in the app when the child or parent uses it to play with the toy, as shown in Fig.~\ref{fig:chip_change_name}.

The toy and all its companion components (Smart Bed, Smart Ball, Smart Band) use static Bluetooth MAC addresses that allow child tracking, leaving it vulnerable to \emph{\bluetooth}. It is likewise vulnerable to \emph{\play} since the connection between the companion app and the toy is over unencrypted BLE. Any installed instance of the app can pair with the toy and have access to full toy functionality. A child can play with \chip\ directly without the need for an app, leaving the toy available for Bluetooth pairing. This could be exploited maliciously if an adversary is within Bluetooth range (about 33 metres). Such an adversary could control the toy, navigating it and making it perform activities such as Yoga or Wanna Play that could injure a child, especially taking into account the weight of the toy at 2.7 kg. An adversary could also make the toy bark continuously by adjusting the wake up alarm, possibly scaring the child.

\subsection{\cloudpets}
\cloudpets\ allows exchanging voice messages between children, parents, and friends. The toy does not connect to the Internet directly, but connects to a companion app through Bluetooth. A parent uses the app to create parent and child accounts, and links the toy to the child's account. The toy receives a link request, appearing as red heart pulses; pressing a button on the toy's hand accepts the connection. Through the parent account, parents can accept or reject messages sent to or from the toy.
\subhead{\pii} 
The \cloudpets\ companion app connects to the server parse-cloudpets.spiraltoys.com and transmits parent and child names, pictures, and dates of birth; parent email address; friend names and profile photos; and child voice messages, which are stored in the parent account in the cloud. It connects to ads and analytics servers which collect PII as follows. Googleads.g.doubleclick.net sets DSID and IDE cookies, which link user activity and feature targeted ads across different platforms~\cite{googlecookies}.
Ads.mopub.com collects the unique device identifier and sends it after hashing using SHA-1. The app also connects to googleadservices.com and googlesyndication.com, which initiate requests to googleads.g.doubleclick.net. The \cloudpets\ companion app sends these unique identifiers even if the user disables ad tracking. 
\balance
\subhead{\secmes}
The toy server suffers from \emph{\tls}. It supports only TLS 1.0, and several anonymous cipher suites (e.g. TLS\_ECDH \_anon\_WITH\_RC4\_128\_SHA) which allow accessing the server without authentication and expose the app-server connection to the risk of MitM attacks. Moreover, the server is vulnerable to RC4, and uses weak key exchange and common DH primes, increasing the risk of losing forward secrecy for transmitted PII. 

PII stored remotely, including audio files and photos, cannot be accessed without authentication. All PII is transmitted over encrypted channel; however, the server does not force using HTTPS. Flipping HTTPS to HTTP sends PII in plaintext, exposing the toy to the \emph{\comm} vulnerability. The toy suffers from \emph{\bluetooth} as it uses a static MAC address and it does not apply BLE privacy; a MitM could access communication between the app and the toy. 

The app displays ads that cover the entire screen at app launch and periodically during use; some of them force users to wait for several seconds before allowing them to close the ads. Moreover, the app features a persistent ad bar at the bottom or top of the interface which continuously displays ads to the user. While many free standalone smartphone apps are monetized through ads, it is unusual to observe aggressive (or any) ad display in the companion app of a purchased toy, and \cloudpets\ is unique in this regard among the toys we examine.

\end{document}